\documentclass[twocolumn,showpacs,preprintnumbers,amsmath,amssymb]{revtex4}
\usepackage{graphicx}% Include figure files
\usepackage{dcolumn}% Align table columns on decimal point
\usepackage{bm}% bold math

\begin{document}

\preprint{APS/123-QED}

\title{Geometrical Representation of Sum Frequency Generation and \\Adiabatic Frequency Conversion}

\author{Haim Suchowski}
\email{Haim.suchowski@weizmann.ac.il}
\author{Dan Oron}
\author{Yaron Silberberg}
\affiliation{Department of Physics of Complex System, Weizmann
Institute of Science, Rehovot, 76100, Israel}
\author{Ady Arie}
\affiliation{School of Electrical Engineeing, Faculty of
Engineering, Tel Aviv University, Tel Aviv, Israel}

\date{\today}

\begin{abstract}
We present a geometrical representation of sum frequency generation
process in the undepleted pump approximation. The analogy of such
dynamics with the known optical Bloch equations is discussed. We use
this analogy to present a novel technique for the achievement of
both high efficiency and large bandwidth in a sum frequency
conversion processes using adiabatic inversion scheme, adapted from
NMR and light-matter interaction. The adiabatic constraints are
derived in this context. Last, this adiabatic frequency conversion
scheme is realized experimentally by a proper design of adiabatic
aperiodically poled KTP device, using quasi phased matched method.
In the experiments we achieved high efficiency signal to idler
conversion over a bandwidth of 140nm.

\end{abstract}
 \maketitle

The generation of tunable frequency optical radiation typically
relies on nonlinear frequency conversion in crystals. In this
process, light of two frequencies is mixed in a nonlinear crystal,
resulting in the generation of a third color with their sum or
difference frequency. These three-wave mixing processes, also known
as frequency up-conversion or frequency down-conversion are
typically very sensitive to the incoming frequencies, owing to lack
of phase matching of the propagating waves. Thus, angle, temperature
or other tuning mechanisms are needed to support efficient frequency
conversion. This difficulty is of particular importance when trying
to efficiently convert broadband optical signals, since simultaneous
phase matching of a broad frequency range is difficult.

Solving the general form of the wave equations governing three wave
mixing processes in nonlinear process is not an easy task. Under
certain conditions these can be simplified to three nonlinear
coupled equations. Further simplification can be applied when one
incoming wave (termed pump) is much stronger than the other two. In
the "undepleted pump" approximation, two linear coupled equation are
obtained rather than three nonlinear ones \cite{Boyd2005}. In the
case of sum frequency generation (SFG) process, this simplified
system possesses SU(2) symmetry, sharing its dynamical behavior with
other two states systems, such as nuclear magnetic resonance (NMR)
or the interaction of coherent light with a two-level atom.
 In this letter we explore the dynamical symmetry of SFG
process in analogy with the well known two level system dynamics
\cite{Eberly1975}. We also apply a geometrical visualization using
the approach presented by Bloch \cite{Bloch1946} and Feynman et al.
\cite{Feynman1957} in NMR and light-matter interaction,
respectively. The simple vector form of the coupling equation allows
for new physical insight into the problem of frequency conversion,
enabling a more intuitive understanding of the effects of spatially
varying coupling and phase mismatch. The utility of this approach is
demonstrated by introducing a robust, highly efficient broadband
color conversion scheme, based on an equivalent mechanism for
achieving full population inversion in atom-photon interactions
known as Rapid Adiabatic Passage (RAP) scheme \cite{Massiah1961}.
The demonstration is experimentally realized using an adiabatically
varying aperiodically poled KTP (APPKTP) in a quasi phased matched
(QPM) design.

Aperiodically poled structure have already been introduced for
improving the bandwidth response of frequency conversion, but at a
cost of a significantly reduced efficiency
\cite{Fejer1994,Mizuuchi1994,Zhu2005}. The broad bandwidth response
is in particular importance for frequency conversion of ultrashort
pulses. Chirped QPM gratings have been utilized to manipulate short
pulses both in second harmonic generation (SHG)
\cite{Arbore1997,Imeshev2000,fejerReview2007}, difference frequency
generation (DFG) \cite{Imeshev2001}, and in parametric amplification
\cite{Imeshev2000, Lefort2008}. Recent application showed that by
using a disordered material (random QPM), an extreme broad bandwidth
was obtained, although at a price of severe reduction of the
conversion efficiencies \cite{Baudrier2004}. While in standard
frequency conversion processes, the crystal parameters and the pump
intensity should be precisely matched in order to reach high
conversion efficiency, here, by utilizing
adiabatic frequency conversion %using these tight tolerances are lifted, and
one can still reach near to $100\%$ conversion efficiency over a
broad wavelength and temperature range. In our demonstration we
succeed in achieving near complete conversion while maintaining
extremely broad bandwidth of over 140nm.

\begin{figure}\label{phasematched}
\includegraphics[width=1\columnwidth]{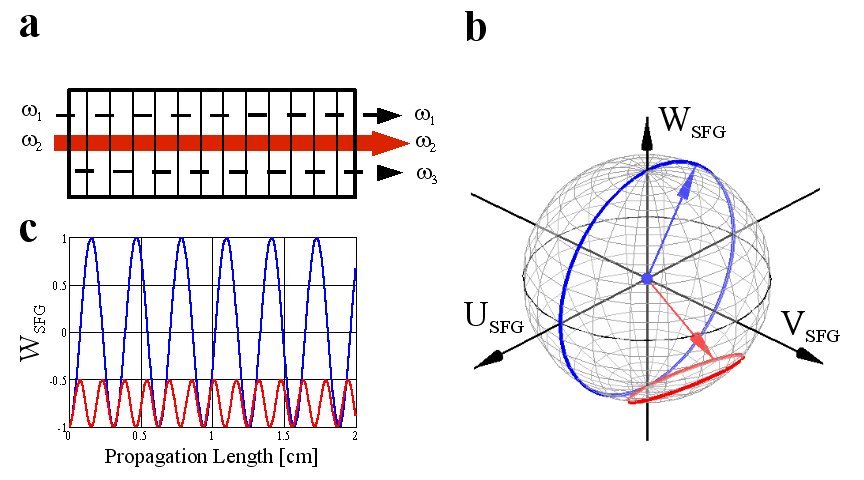}
\caption{\small Bloch sphere geometrical representation of SFG in
the undepeleted pump approximation. The amplitude of $\omega_{2}$
(bold red arrow) is much larger than the amplitudes of $\omega_{1}$,
$\omega_{3}$, and therefore remains constant along the propagation.
The energy transfer between the incoming weak field ($\omega_{1}$)
and the converted SFG signal($\omega_{3}$) along the propagation is
dictated by the coupling coefficient $\kappa$ and the phase mismatch
$\Delta{k}$. a) A periodically poled quasi phased crystal designed
to achieve a perfect phase matched interaction along the
propagation. (b) Geometrical visualization of the SFG dynamics on a
SFG Bloch Sphere. Two trajectories are plotted: perfect phase
matching (blue, torque vector points to the equator) which can
result in efficient conversion and a constant nonzero phase mismatch
(red, torque vector points to a point in the south hemisphere),
\emph{always} resulting in an inefficient conversion process. (c)
The projection of the trajectory onto the z axis yields the
conversion efficiency.}
\end{figure}

Let us first consider the geometrical representation of SFG. In the
undepleted pump approximation, the pump amplitude is assumed
constant along the nonlinear crystal, and the following normalized
coupled equations for the signal and idler can be constructed
\cite{Boyd2005}:
\begin{subequations}
\begin{eqnarray}
i\frac{d\tilde{A_{1}}}{dz}&=&\kappa\tilde{A_{3}}e^{-i\Delta{k}z}\\
i\frac{d\tilde{A_{3}}}{dz}&=&\kappa^{*}\tilde{A_{1}}e^{+i\Delta{k}z}
\end{eqnarray}\label{coupled}
\end{subequations}
where $\Delta{k}=k_{1}+k_{2}-k_{3}$ is the phase mismatch, z is the
position along the propagation axis, $\kappa$ is the coupling
coefficient, and $\tilde{A_{1}}$ and $\tilde{A_{3}}$ are the
normalized signal and idler amplitudes. The coupling coefficient and
the normalized amplitudes are written below using the following
quantities:
\begin{subequations}
\begin{eqnarray}
K_{1}&=&\frac{16\pi{i}\omega_{1}^{2}\chi^{(2)}}{k_{1}c^{2}}A_{2}^{*},\:\:K_{3}=\frac{16\pi{i}\omega_{3}^{2}\chi^{(2)}}{k_{3}c^{2}}A_{2}\\
\kappa&=&{i}\sqrt{K_{1}K_{3}}\\
\tilde{A_{1}}&=&A_{1}/\sqrt{K_{1}},\:\:\tilde{A_{3}}=A_{3}/\sqrt{K_{3}}
\end{eqnarray}\label{Boyd_def}
\end{subequations}
where $\omega_{1}$ and $\omega_{3}$ are the frequencies of the
signal and idler, respectively, $k_{1}$ and $k_{3}$ are their
associated wave numbers, $c$ is the speed of light in vacuum,
$A_{1}$, $A_{2}$, $A_{3}$ are the signal, pump and idler amplitudes
respectively, and $\chi^{\left(2\right)}$ is the $2^{nd}$ order
susceptibility of the crystal (assumed to be frequency independent).

The dynamics of this system is dictated by the values of $\Delta{k}$
and $\kappa$. Full energy transfer from signal to idler (SFG
process) or vice versa (DFG process) is usually achieved by
demanding perfect phase matching $\left(\Delta{k}(z)=0\right)$, and
also by satisfying a constraint relating the propagation length in
the crystal with the pump intensity, such that $\kappa\cdot
z=\left(2n+1\right)\pi$, where n is an integer. This is the only
solution for a full conversion with a z-independent phase mismatch.

\begin{table*}
\caption{\small Analogy between Sum Frequency Generation (SFG)
process in the undepleted pump approximation and the dynamics of a
two level atomic system, induced by coherent light. The middle
column describes the optical Bloch sphere realization of the two
level systems (notation are taken from Allen $\&$ Eberly
\cite{Eberly1975}). The right column shows the analogous parameters
of the SFG sphere realization.}
\begin{tabular}{|c|c|c|}%\label{tab}
  \hline
  Parameter & Optical Bloch Sphere & SFG Sphere Realization \\
  \hline
  Evolution parameter & time & z axis \\
  Ground state population & $|a_{g}|^{2}$ & $|A_{1}|^{2}$ \\
  Excited state population & $|a_{e}|^{2}$ & $|A_{3}|^{2}$ \\
  Energy difference & $\omega_{0}=\omega_{fg}$ & $n\left(\omega_{2}\right)\omega_{2}/{c}$\\
  Detuning / Phase Mismatch & $\Delta$ & $\Delta{k}$\\
  "Rabi" frequency & $\Omega_{0}=\frac{1}{\hbar}\mu\cdot{E_{in}}$ & $\kappa=\frac{4\pi
w_{1}w_{3}}{(k_{1}k_{3})^{1/2}c^{2}}\chi^{(2)}\cdot{E_{2}}$\\
Torque vector &
$\Omega=\left(Re\{\Omega_{0}\},Im\{\Omega_{0}\},\Delta\right)$ &
$g=\left(Re\{\kappa\},Im\{\kappa\},\Delta{k}\right)$\\
State vector - $\rho_{X}$=U & $a_{f}^{*}a_{g}+a_{g}^{*}a_{f}$ & $A_{3}^{*}A_{1}+A_{1}^{*}A_{3}$\\
State vector - $\rho_{Y}$=V & $i(a_{f}^{*}a_{g}-a_{g}^{*}a_{f})$ & $i(A_{3}^{*}A_{1}-A_{1}^{*}A_{3})$\\
State vector - $\rho_{Z}$=W & $|a_{f}|^2-|a_{g}|^2$ & $|A_{3}|^2-|A_{1}|^2$\\
  \hline
\end{tabular}
\end{table*}

Equations \ref{coupled}a \& \ref{coupled}b have the same form as
those describing the dynamics in two level systems. We adopt Feynman
et al. \cite{Feynman1957} formulation and write the dynamics of this
problem as a real three dimensional vector equation, which can be
visualized geometrically on a sphere, known as Bloch sphere. In this
context, we define a three dimensional state vector
$\vec{\rho_{SFG}}=\left(U,V,W\right)$ as follows:
\begin{subequations}
\begin{eqnarray}
U_{SFG} &=& A_{3}^{*}A_{1}+A_{1}^{*}A_{3}\:,\\
V_{SFG} &=& i(A_{3}^{*}A_{1}-A_{1}^{*}A_{3})\:,\\
W_{SFG} &=& |A_{3}|^2-|A_{1}|^2\:.
\end{eqnarray}
\end{subequations}

This vector represents the relation between the signal and idler
along the crystal, and includes the coherence between the modes
along the propagation. In particular, its z component ($W_{SFG}$)
gives information about the conversion efficiency. The south pole
$\vec{\rho}=(0,0,-1)$ corresponds to zero conversion ($A_{3}=0$),
while the north pole $\vec{\rho}=(0,0,1)$ corresponds to full
conversion. In between, the conversion efficiency can be found as
follow: $\eta=\left(W_{SFG}+1\right)/2$. The torque vector
$\vec{g}=\left(Re\{\kappa\},Im\{\kappa\},\Delta{k}\right)$
represents the coupling between these two modes, which contains also
the phase mismatch parameter. In this form, the loss-free evolution
equations can be written as a single vector precession equation:
\begin{equation}
\large\frac{d\vec{\rho}_{\small{SFG}}}{dz}=\vec{g}\times\vec{\rho}_{\small{SFG}}
\end{equation}

The analogy suggests the equivalence between the quantities in the
standard two-level system and those in the frequency-conversion
problem: The population of the ground and excited states are
analogous to the magnitude of the signal and idler fields ,
respectively. Time evolution is replaced by propagation in the
longitudinal z-axis, and the detuning $\Delta$ is replaced by the
phase-mismatch $\Delta k$ value; The analogy is further detailed in
table I.

Presented in this context, the perfect phased matched solution for
full conversion, as described above, has the same dynamical
trajectory on the Bloch sphere surface as on-resonant light-matter
interaction in atomic physics. This results in oscillatory dynamics
between the two modes ("Rabi oscillations"). The case of an odd
$\pi$-pulse is analogous to transferring the energy from
$\omega_{1}$ (input signal) to the SFG/DFG output $\omega_{3}$, i.e.
the SFG state vector is rotated from the south pole (north pole) to
the north pole (south pole). Any phase mismatch will lead to a
dynamics similar to a detuned case in light-matter interactions,
which exhibit faster oscillations and lower conversion efficiencies,
as shown in figure 1\ref{phasematched}.

\begin{figure}
\includegraphics[width=1\columnwidth]{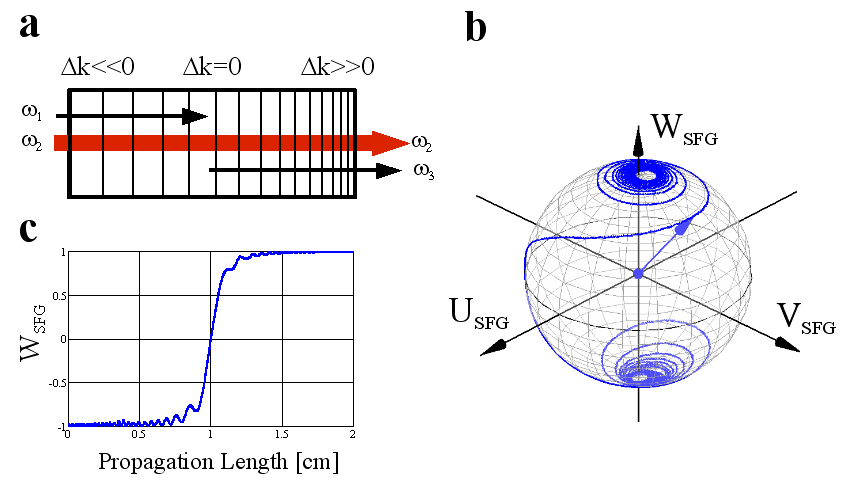}
\caption{\small Adiabatic conversion scheme of SFG in the undepleted
pump approximation. a) Continuous adiabatic variation of the phase
mismatch parameter is required. This can be achieved by slowly
changing the poling periodicity along the propagation direction. b)
The adiabatic following trajectory, where the torque vector
initially points to the vicinity of the south pole, and ends up at
the vicinity of the north pole. c) The projection of the trajectory
onto the z axis yields the conversion efficiency. In this
trajectory, phase matching condition is fulfilled at
z=1cm.}\label{Adiabatic}
\end{figure}

Due to the above, achievement of full energy transfer in an
upconversion process is usually not robust, requiring several
ingredients to be simultaneously satisfied. From the above analogy
we can, however, adopt the RAP mechanism \cite{Massiah1961}, where a
strong chirped excitation pulse scans slowly through the resonance
to achieve robust full inversion, to the realm of frequency
conversion. Thus, in order to transfer from $A_{1}(z)$ to $A_{3}(z)$
for a broadband of wavelengths, the phase mismatch parameter,
$\Delta{ k}\left(z\right)$ has to change adiabatically from a big
negative value to a big positive value (or vice versa). The
adiabaticity conditions require that:
\begin{subequations}
\begin{eqnarray}
|\Delta{k}|&>>&\kappa\\
\Delta{k}(z=0)<0&, \:&\:\Delta{k}(z=L)>0 \\
\left|\frac{d\Delta{k}}{dz}\right|&<<&
\frac{\left(\Delta{k}^{2}+\kappa^{2}\right)^{3/2}}{\kappa}
\end{eqnarray}\label{adiabatic_eq}
\end{subequations}

The first two conditions deal with the maximal magnitude of the
phase mismatch, which should be very large in comparison to the
value of $\kappa$, and the interaction has to start with a large
negative (or positive) value of phase mismatch, and end with a large
positive (or negative) value. The third condition is the most acute
one, dealing with the rate of change of the phase mismatch during
propagation. To satisfy this condition, the phase mismatch has to be
vary slowly as compared to the square of the coupling term in this
nonlinear process. If the rate of variation is not slow enough (left
hand side value of inequality \ref{adiabatic_eq}c is not small
enough), or the coupling coefficient is not large enough, this
inequality won't be satisfied and the conversion efficiency will be
poor. Figure \ref{Adiabatic} demonstrate the case when all those
constraints are satisfied and full frequency conversion is achieved.

In the experimental realization of this adiabatic conversion scheme
we choose to utilize the QPM technique \cite{Armstrong1963}. By
tuning the spatial structure of the domains, this technique allows
not only to achieve phase mismatch that is close to zero along the
propagation axis, but almost any desired function of the phase
mismatched parameter. By proper design of the periodicity of the
poling, one can achieve an effective phase mismatch parameter which
is the summation of the process phase mismatch and the artificial
phase-mismatch $\Delta{k_{\Lambda}\left(z\right)}$:

\begin{equation}
\Delta{k_{eff}\left(z\right)}=k_{signal}+k_{pump}-k_{idler}-\Delta{k_{\Lambda}\left(z\right)}\:.
\end{equation}
The desired value of the phase-mismatch parameter is achieved by
poling the QPM structure using the approximate relation:
$\Delta{k_{\Lambda}\left(z\right)}=\frac{2\pi}{\Lambda\left(z\right)}\:$,
where $\Lambda\left(z\right)$ is the local poling period.

We demonstrated our idea by an adiabatic APPKTP which was designed
 to satisfy the constraints posed by equations \ref{adiabatic_eq}.
The periodicity was varied from 14.6 $\mu{m}$ to 16.2 $\mu{m}$ along
a crystal length of $L=20mm$, to induce a linear adiabatic change in
$k_{eff}\left(z\right)$. The design was carried out by a numerical
simulations of the process propagation using the finite difference
method. We assumed plain wave approximation and took
$\chi^{\left(2\right)}=32\frac{pm}{V}$. We used an OPO (Ekspla
NT342) as our sources which produced a strong pump at 1064 nm (6 ns,
130 $\mu{J}$), and a tunable signal varied from 1400 nm to 1700 nm
(5 ns, $~$ 1 $\mu{J}$). The pump and the signal, both polarized in
the extra-ordinary axis were spatially overlapped and focused
collinearly into the crystal having waists of $150\mu{m}^{2}$ and
$120\mu{m}^{2}$, respectively. These values guarantee that the
Rayleigh range is larger than the crystal length and thus the plane
wave approximation of our simulations holds. We collected both the
input wavelength (signal) and output SFG wavelength (idler) after
their propagation in the crystal and recorded by an InGaAs detector,
and a cooled CCD spectrometer, respectively.

\begin{figure}
\includegraphics[width=1\columnwidth]{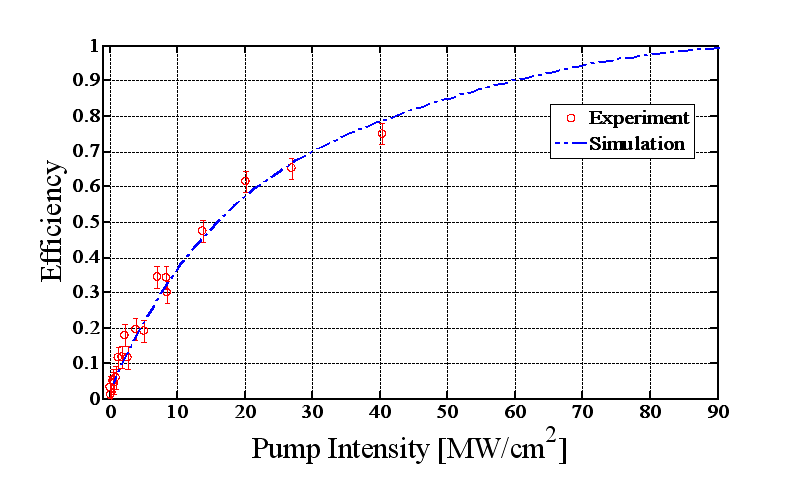}
\caption{\small Conversion efficiency as a function of pump
intensity with a signal wavelength of 1530nm at room temperature
($21^{\circ}{C}$). A good correspondence between the experimental
results (shown in red) and the simulation of the design
(dashed-dotted blue). The maximal conversion efficiency achieved was
$74\%\pm3\%$.%pump intensity larger than 60MW/cm^2 more than 90% of the input photons are transformed to output SFG photons.
}\label{Pump_freq_comp}
\end{figure}

\begin{figure}
\includegraphics[width=1\columnwidth]{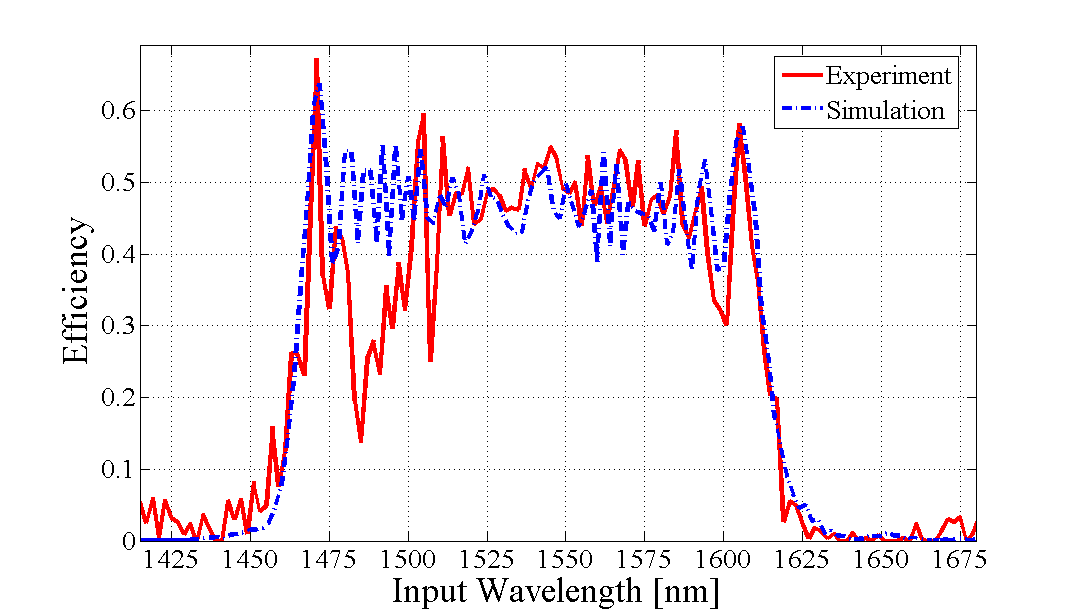}
\caption{\small Conversion efficiency as a function of input
wavelength using the adiabatic APPKTP design at a pump intensity of
15 $MW/cm^{2}$. A good correspondence between the experimental
results (shown in solid red) and the simulation of the design
(dashed-dotted blue) is shown. The low efficiency around 1485nm is
associated with a manufacturing defect.}\label{wavelength_dep}
\end{figure}

Our demonstration consists of two sets of experiments. First, to
demonstrate the crucial requirement of the strong pump, we examined
the dependence of the conversion efficiency on the pump intensity.
In the second experiment we measure the conversion efficiency as a
function of signal wavelength at a constant pump intensity. In both
cases, the conversion efficiency was measured by comparing the
signal intensity with and without the presence of the pump. This was
checked to be completely correlated with the observed SFG intensity
and free of thermal effects.

In Fig. \ref{Pump_freq_comp} we plot the conversion efficiency as a
function of the pump peak intensity for a fixed signal frequency
1530 nm. A very good correspondence is obtained with the simulation.
The maximal efficiency which was achieved with our maximal pump
intensity was $74\%\pm3\%$.

In a second experiment, when the input wavelength was varied, we
chose to work at a moderate pump intensity of 15 $MW/cm^{2}$ and at
room temperature. We show in Figure \ref{wavelength_dep} efficient
broadband conversion of over 140nm wide (1470nm to 1610nm). This is
in a good correspondence with the numerical simulation of the design
(dashed-dotted blue), except for a small region of low efficiency
around 1485nm, which is associated with a manufacturing defect, thus
causing the violation of the adiabaticity condition. For comparison,
a conventional efficient converter (a periodically poled structure
designed to achieve perfect phase matching), would lead to efficient
broadband conversion only for 2nm bandwidth.

Other experiments showing the robustness of the adiabatic conversion
scheme to errors in incidence angle, crystal temperature and pump
frequency (as compared to the periodically poled design), and
exhibiting color tunability of the conversion band by temperature
control will be presented in a forthcoming publication.

In conclusion, we show that the use of a geometrical Bloch sphere
visualization to describe evolution of the complex mode amplitudes,
gives information on their relative phases as well as intensities.
We demonstrated high efficiency wavelength conversion for a wide
range of frequencies, by using the adiabatic conversion method
adapted from other well known two level systems. It is important to
note that quasi phase matched crystal is only one possible
realization; the same mechanism can be applied, for example, by
inducing a temperature gradient on a perfect phased matched crystal,
or by any other mechanism fulfilling the adiabatic constraint of Eq.
\ref{adiabatic_eq}. The present scheme can be utilized for efficient
frequency conversion of broadband signals as well as ultrashort
pulses. This analysis holds promise for ultrashort pulse conversion
in a wide range of frequencies from UV to far IR. It may be
particularly useful in the efficient upconversion of weak infrared
signals to the near-infrared or visible range, used in atmospherical
observation \cite{Brustlein2008,Boyd1977}. The generalization of
this analogy to higher order nonlinear processes interaction can be
made, and known schemes from
atom-photon interaction or NMR, such as %EIT \cite{Harris1997},
Stimulated Raman Adiabatic Passage (STIRAP) \cite{Vitanov2001} can
be adopted.

This research was supported by Israel Science Foundation (Grants No.
1621/07 and 960/05). H. S. is grateful to the Azrieli Foundation for
the award of an Azrieli Fellowship.

\end{document}